\documentclass[conference, hidelinks]{IEEEtran}

\usepackage{amsmath,amssymb,amsfonts}
\usepackage{array}
\usepackage[caption=false,font=normalsize,labelfont=sf,textfont=sf]{subfig}
\usepackage{textcomp}
\usepackage{stfloats}
\usepackage{url}
\usepackage{verbatim}
\usepackage{graphicx}
\usepackage{cite}
\usepackage{xcolor}
\usepackage{orcidlink}
\usepackage[acronym,shortcuts]{glossaries}
\usepackage{bm}
\usepackage{relsize}
\usepackage{soul}
\usepackage{algorithm}
\usepackage[noend]{algpseudocode}
\usepackage{algcompatible}
\usepackage{mathtools}
\usepackage{bm}
\usepackage{lipsum}
\usepackage{mathtools}
\usepackage{lipsum}
\usepackage[bottom]{footmisc}
\usepackage{tabularx}
\def\BibTeX{{\rm B\kern-.05em{\sc i\kern-.025em b}\kern-.08em
T\kern-.1667em\lower.7ex\hbox{E}\kern-.125emX}}

\newcommand{\trans}[0]{^{\mathsf{T}}}

\newacronym{PCA}{PCA}{principal component analysis}
\newacronym{MDSM}{MDSM}{multi-domain sparse modulation}
\newacronym{P2P}{P2P}{point-to-point}
\newacronym{OTAC}{AirComp}{over-the-air computing}
\newacronym{TX}{TX}{transmitter}
\newacronym{RX}{RX}{receiver}
\newacronym{IoT}{IoT}{Internet-of-things}
\newacronym{AI/ML}{AI/ML}{artifitial intelligence/machine learning}
\newacronym{SDR}{SDR}{semi-definite relaxation}
\newacronym{EVD}{EVD}{eigenvalue decomposition}
\newacronym{GR}{GR}{Gaussian randomization}
\newacronym{SCA}{SCA}{successive convex approximation}
\newacronym{BnB}{BnB}{branch and bound}
\newacronym{QT}{QT}{quadratic transform}
\newacronym{RQ}{RQ}{Rayleigh quotient}
\newacronym{SOCP}{SOCP}{second-order cone programming}
\newacronym{CDF}{CDF}{cumulative distribution function}
\newacronym{UF}{UF}{uniform-forcing}
\newacronym{AP}{AP}{access point}
\newacronym{RSDR}{R-SDR}{regularized semi-definite relaxation}
\newacronym{R-SDR}{R-SDR}{regularized SDR}
\newacronym{flops}{flops}{floating point operations}
\newacronym{ED}{ED}{edge device}
\newacronym{SINR}{SINR}{signal to interference-plus-noise ratio}
\newacronym{SIC}{SIC}{successive interference cancellation}
\newacronym{CSI}{CSI}{channel state information}
\newacronym{LoS}{LoS}{line-of-sight}
\newacronym{NLoS}{NLoS}{non-LoS}
\newacronym{RPE}{RPE}{radar parameter estimation}
\newacronym{OTFS}{OTFS}{orthogonal time frequency space}
\newacronym{AFDM}{AFDM}{affine frequency division multiplexing}
\newacronym{CRLB}{CRLB}{Cram{\`e}r-Rao lower bound}
\newacronym{BCRLB}{BCRLB}{Bayesian Cram{\`e}r-Rao lower bound}
\newacronym{BBI}{BBI}{Bayesian bilinear inference}
\newacronym{AoA}{AoA}{angle-of-arrival}
\newacronym{SNR}{SNR}{signal-to-noise ratio}
\newacronym{ML}{ML}{maximum likelihood}
\newacronym{MIMO}{MIMO}{multiple-input multiple-output}
\newacronym{SIMO}{SIMO}{single-input multiple-output}
\newacronym{SISO}{SISO}{single-input single-output}
\newacronym{MUSIC}{MUSIC}{multiple signal classification}
\newacronym{MU}{MU}{multi-user}
\newacronym{ROOT-MUSIC}{ROOT-MUSIC}{ROOT multiple signal classification}
\newacronym{JCAS}{JCAS}{joint communication and sensing}
\newacronym{JCR}{JCR}{joint communications and radar}
\newacronym{ISAC}{ISAC}{integrated sensing and communications}
\newacronym{3D}{3D}{three-dimensional}
\newacronym{2D}{2D}{two-dimensional}
\newacronym{1D}{1D}{one-dimensional}
\newacronym{BF}{BF}{beamforming}
\newacronym{ROI}{ROI}{region of interest}
\newacronym{mmWave}{mmWave}{millimeter-wave}
\newacronym{MF}{MF}{matched-filter}
\newacronym{DD}{DD}{delay-Doppler}
\newacronym{SotA}{SotA}{state-of-the-art}
\newacronym{ULA}{ULA}{uniform linear array}
\newacronym{QAM}{QAM}{quadrature amplitude modulation}
\newacronym{ISFFT}{ISFFT}{inverse symplectic finite Fourier transform}
\newacronym{SFFT}{SFFT}{symplectic finite Fourier transform}
\newacronym{ISI}{ISI}{inter-symbol interference}
\newacronym{AWGN}{AWGN}{additive white Gaussian noise}
\newacronym{MSE}{MSE}{mean-squared-error}
\newacronym{LMMSE}{LMMSE}{linear minimum mean square error}
\newacronym{RMSE}{RMSE}{root mean square error}
\newacronym{ESPRIT}{ESPRIT}{estimation of signal parameters via rotational invariant techniques}
\newacronym{OFDM}{OFDM}{orthogonal frequency division multiplexing}
\newacronym{OCDM}{OCDM}{orthogonal chirp division multiplexing}
\newacronym{BS}{BS}{base station}
\newacronym{UE}{UE}{user equipment}
\newacronym{JCEDD}{JCEDD}{joint channel estimation and data detection}
\newacronym{PDA}{PDA}{probabilistic data association}
\newacronym{PMF}{PMF}{probability mass function}
\newacronym{PBiGaBP}{PBiGaBP}{parametric bilinear Gaussian belief propagation}
\newacronym{PBiGAMP}{PBiGAMP}{parametric bilinear generalized approximate message passing}
\newacronym{GaBP}{GaBP}{Gaussian belief propagation}
\newacronym{FT}{FT}{frequency-time}
\newacronym{DFT}{DFT}{discrete Fourier transform}
\newacronym{IDFT}{IDFT}{inverse discrete Fourier transform}
\newacronym{TD}{TD}{time domain}
\newacronym{wlg}{w.l.g.}{without loss of generality}
\newacronym{CP}{CP}{cyclic prefix}
\newacronym{DAF}{DAF}{discrete affine Fourier}
\newacronym{DAFT}{DAFT}{discrete affine Fourier transform}
\newacronym{IDAFT}{IDAFT}{inverse discrete affine Fourier transform}
\newacronym{CPP}{CPP}{\textit{chirp-periodic} prefix}
\newacronym{IDZT}{IDZT}{inverse discrete Zak transform}
\newacronym{DZT}{DZT}{discrete Zak transform}
\newacronym{P/S}{P/S}{parallel-to-serial}
\newacronym{S/P}{S/P}{serial-to-parallel}
\newacronym{SBL}{SBL}{sparse Bayesian learning}
\newacronym{MPA}{MPA}{message passing algorithms}
\newacronym{EM}{EM}{expectation maximization}
\newacronym{sIC}{soft IC}{soft interference cancellation}
\newacronym{soft RG}{soft RG}{soft replica generation}
\newacronym{BG}{BG}{belief generation}
\newacronym{SGA}{SGA}{scalar Gaussian approximation}
\newacronym{CLT}{CLT}{central limit theorem}
\newacronym{PDF}{PDF}{probability density function}
\newacronym{QPSK}{QPSK}{quadrature phase-shift keying}
\newacronym{ICI}{ICI}{inter-carrier interference}
\newacronym{BER}{BER}{bit error rate}
\newacronym{DoF}{DoF}{degrees-of-freedom}
\newacronym{VGA}{VGA}{vector Gaussian approximation}
\newacronym{FD}{FD}{full-duplex}
\newacronym{NMSE}{NMSE}{normalized mean square error}
\newacronym{KL}{KL}{Kullback-Leibler}
\newacronym{LASSO}{LASSO}{least absolute shrinkage and selection operator}
\newacronym{FP}{FP}{fractional programming}
\newacronym{CC}{CC}{communication-centric}
\newacronym{RC}{RC}{raised-cosine}
\newacronym{RRC}{RRC}{root raised-cosine}
\newacronym{6G}{6G}{sixth-generation}
\newacronym{V2X}{V2X}{vehicle-to-everything}
\newacronym{LEO}{LEO}{low-earth orbit}
\newacronym{I/O}{I/O}{input-output}
\newacronym{CE}{CE}{channel estimation}
\newacronym{ICC}{ICC}{integrated communication and computing}
\newacronym{ISCC}{ISCC}{integrated sensing, communications and computing}
\newacronym{PAM}{PAM}{pulse amplitude modulation}
\newacronym{iid}{i.i.d.}{independent and identically distributed}
\newacronym{FH}{FH}{frequency-hopping}
\newacronym{JRC}{JRC}{joint radar and communication}
\newacronym{DFRC}{DFRC}{dual-function radar communications}
\newacronym{AF}{AF}{ambiguity function}
\newacronym{PRI}{PRI}{pulse repetition interval}
\newacronym{PRF}{PRF}{pulse repetition frequency}
\newacronym{PPH}{PPH}{polynomial-phase hopping}
\newacronym{QSM}{QSM}{quadrature spatial modulation}
\newacronym{CPM}{CPM}{continuous phase modulation}
\newacronym{ECCM}{ECCM}{electronic counter-countermeasures}
\newacronym{RFA}{RFA}{random frequency agility}
\newacronym{RPA}{RPA}{random pulse repetition interval (PRI) agility}
\newacronym{RFPA}{RFPA}{random frequency and PRI agility}
\newacronym{ASK}{ASK}{amplitude shift keying}
\newacronym{PSK}{PSK}{phase shift keying}
\newacronym{RIP}{RIP}{restricted isometry property}
\newacronym{OMP}{OMP}{orthogonal matching pursuit}
\newacronym{CIR}{CIR}{channel impulse response}
\newacronym{TDD}{TDD}{time division duplex}
\newacronym{CRKG}{CRKG}{channel reciprocity-based key generation}
\newacronym{FCM}{FCM}{Fuzzy C-means}
\newacronym{PIRS}{PIRS}{polar code-based information scheme}
\newacronym{CRC}{CRC}{cyclic redundancy check}
\newacronym{BDR}{BDR}{bit disagreement rate}
\newacronym{SQ}{SQ}{scalar quantization}
\newacronym{VQ}{VQ}{vector quantization}
\newacronym{AI}{AI}{artificial intelligence}
\newacronym{V2V}{V2V}{vehicle to vehicle}
\newacronym{IM}{IM}{index modulation}
\newacronym{SM}{SM}{spatial modulation}
\newacronym{SIM}{SIM}{spatial index modulation}
\newacronym{PLS}{PLS}{physical layer security}
\newacronym{CU}{CU}{communication user}
\newacronym{QoS}{QoS}{quality-of-service}
\newacronym{MMSE}{MMSE}{minimum mean squared error}
\newacronym{CRB}{CRB}{Cramér-Rao bound}
\newacronym{DL}{DL}{deep learning}
\newacronym{RIS}{RIS}{reconfigurable intelligent surfaces}
\newacronym{THz}{THz}{terahertz}
\newacronym{JFI}{JFI}{Jain's fairness index}
\newacronym{AN}{AN}{artificial noise}
\newacronym{LMI}{LMI}{linear matrix inequality}

\begin{document}

\title{Secrecy-Driven Beamforming for Multi-User Integrated Sensing and Communication}

\author{
Ali Khandan Boroujeni$^{*\dagger}$,
Hyeon Seok Rou$^{\ddagger}$,
Ghazal Bagheri$^{\dagger}$,
Kuranage Roche Rayan Ranasinghe$^{\ddagger}$,\\
Giuseppe Thadeu Freitas de Abreu$^{\ddagger}$,
Stefan K\"{o}psell$^{*\dagger}$,
and Rafael F. Schaefer$^{*\dagger}$ \\[1.5ex]
\textit{\small $^{*}$Barkhausen Institut, 01067 Dresden, Germany} \\[-0.75ex]
\textit{\small $^{\dagger}$Technische Universit\"{a}t Dresden, 01069 Dresden, Germany} \\[-0.75ex]
\textit{\small $^{\ddagger}$School of Computer Science and Engineering, Constructor University, 28759 Bremen, Germany}%

\vspace{-2ex}
}

\maketitle
\setcounter{footnote}{1}
\footnotetext{\emph{An extended version of this work has been submitted to the IEEE Journal on Selected Areas in Communications\cite{Boroujeni_arxiv25}, while this paper has distinct contributions.}}

\begin{abstract}
This paper proposes a secure \ac{ISAC} framework for multi-user systems with multiple \acp{CU} and adversarial targets, where the design problem is formulated to maximize secrecy rate under joint sensing and communication constraints. 
An efficient solution is presented based on an accelerated fractional programming method using a non-homogeneous complex \ac{QT}, which decomposes the problem into tractable subproblems for beamforming and \ac{AN} optimization. 
Unlike conventional artificial noise strategies, the proposed approach also exploits \ac{AN} to enhance sensing while avoiding interference with legitimate users. 
Simulation results show significant gains in secrecy rate, communication reliability, and sensing accuracy, confirming the effectiveness and scalability of the proposed framework.
\end{abstract}

\section{Introduction}
\label{sec:introduction}

\glsresetall

A key innovation of \ac{6G} wireless systems over preceding generations will be the integration of sensing and communications functionalities, referred to as \ac{ISAC} \cite{Liu_JSAC22, Lu_IoTJ24, Rou_SPM24}, which addresses the demand for significant efficiency and support for applications such as autonomous driving, smart cities, and \ac{IoT} \cite{ZhangIoTJ2024, WeiCOMMAG2022}.
However, such integration of applications and advanced technologies also introduces challenges in terms of security, interference management, and complexity in \ac{MU} settings \cite{LiIoTJ2025, ZhuoCOMMAG2024}.

Among these challenges, security is particularly critical, as adversaries may exploit shared signals to compromise both communication confidentiality and sensing reliability \cite{SuTWC2021, ChuTVT2023, MitevBOOK2024, BoroujeniJCS_2024}. 
Therefore, the use of common spectrum and hardware also enlarges the attack surface, creating opportunities for adversaries to extract sensitive information or disrupt communication \cite{ZhaoWCL2025, boroujeni2025frequencyhoppingwaveformdesign}. 
To counter such threats, \ac{PLS} methods exploit channel properties to strengthen security without relying solely on cryptography \cite{mitev2023physical, mucchi2021physical, BagheriJCN_2025, BagheriGIIS_2024}, including methods such as robust beamforming, and secure waveform design \cite{DongTGCN2023, PengTVT2025, Rou_WCL25}, which highlight both the potential and the trade-offs between security and performance.

Addressing these trade-offs often leads to non-convex optimization problems shaped by the interplay of sensing, communication, and security requirements \cite{II-ShenTSP2018, ShenJSAC2024}. 
Fractional programming and the Lagrange dual complex \ac{QT} have proven effective in decomposing such problems into tractable subproblems, enabling scalable optimization \cite{ChenWCSP2024, UchimuraTWC2025}, while other approaches such as weighted \ac{MMSE}, ADMM, and SCA \cite{ShiTSP2011, BoydBOOK2011, RazaviyaynBOOK2013} further reduce the computational effort, with recent works combining machine learning techniques to improve the adaptability \cite{Lu_IoTJ24, LiuTWC2025}.

Despite this progress, practical deployment remains constrained by hardware limitations, imperfect \ac{CSI}, and computational complexity, calling for frameworks that are both secure and efficient while remaining scalable to large system dimensions.

Therefore, in this paper, we propose a secure and scalable framework for \ac{MU} \ac{ISAC} systems with multiple \acp{CU} and adversarial targets.
Unlike prior studies limited to single-target models, our formulation explicitly accounts for multiple sensing targets that may also act as eavesdroppers, capturing realistic adversarial conditions.
We maximize secrecy rate under strict sensing and communication requirements via an efficient Lagrangian dual framework with a non-homogeneous complex \ac{QT}, which decomposes the problem into tractable subproblems with closed-form iterative updates.
In contrast to conventional approaches where \ac{AN} serves only as jamming, our design leverages \ac{AN} as a dual-purpose enabler that enhances both secrecy and sensing while remaining confined to the null space of legitimate users, with robust beamforming to handle angular uncertainty.
As a result, the proposed framework achieves provable secrecy gains, adaptability, robustness, and scalability for large-scale \ac{MIMO} \ac{ISAC} in next-generation wireless networks.

The contributions of this paper are summarized as follows:
\begin{itemize}
\item A secure \ac{MU} \ac{ISAC} formulation modeling multiple adversarial targets as potential eavesdroppers.
\item A dual fractional programming framework based on a non-homogeneous \ac{QT}, with closed-form updates avoiding costly matrix inversions.
\item A joint beamforming and \ac{AN} design that transforms \ac{AN} from a pure jammer into a dual-purpose enabler for secrecy and sensing, while ensuring robustness under angular uncertainty and scalability for large-scale \ac{MIMO} \ac{ISAC}.
\end{itemize}

\section{System Model}
\label{sec:sys_model}

\begin{figure}[H]
\centering
\includegraphics[width=0.9\linewidth]{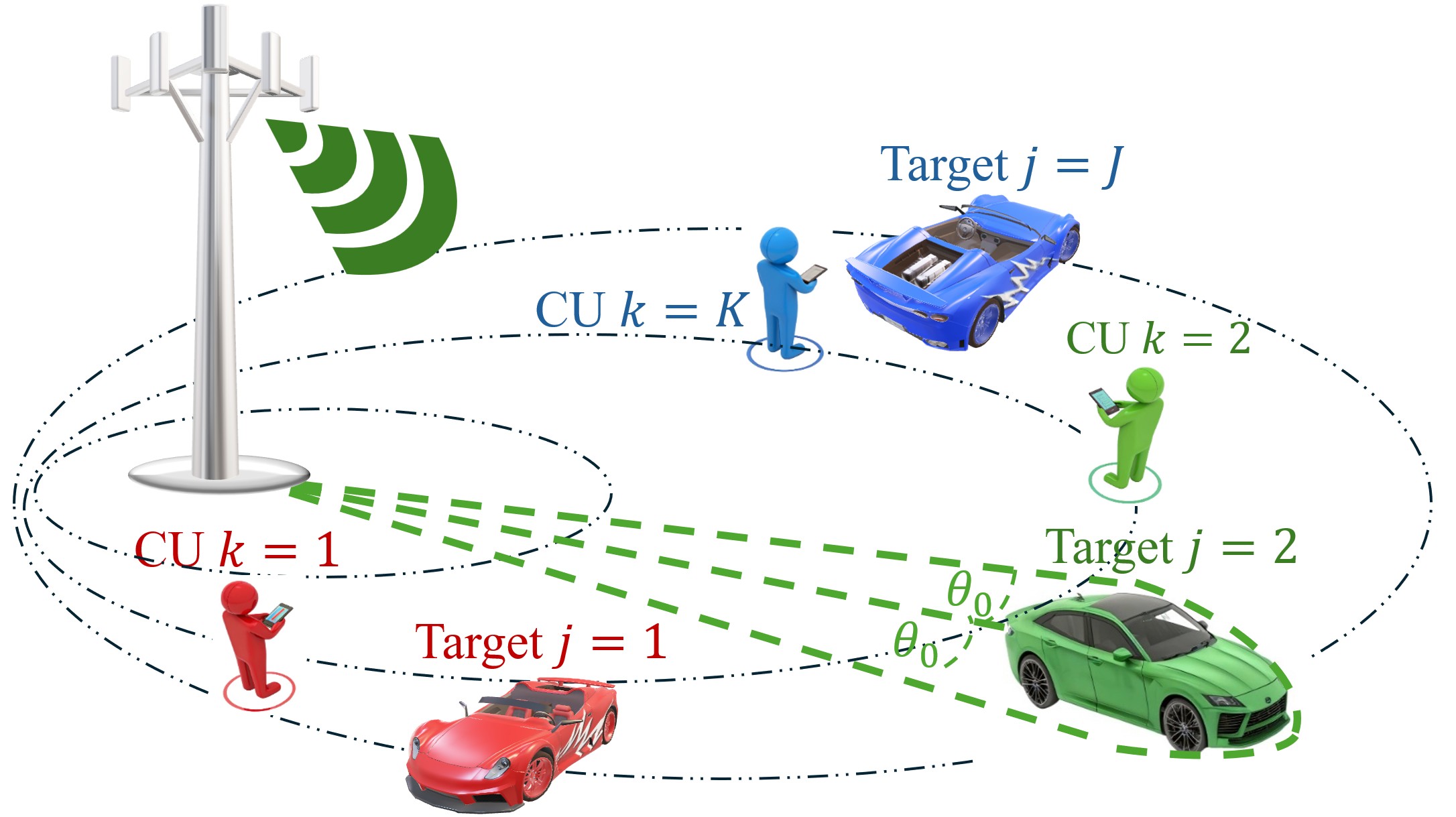}
\vspace{-2ex}
\caption{System model of a secure \ac{MU}-\ac{ISAC} network, where a multi-antenna \ac{BS} serves $K$ \acp{CU} while sensing $J$ targets that may act as eavesdroppers. Each target's direction $\theta_j$ is assumed to be pre-estimated.}
\label{fig:system_model}
\end{figure}

We consider a \ac{MU} \ac{ISAC} system where a base station (\ac{BS}) with $N_t$ transmit antennas serves $K$ single-antenna legitimate \acp{CU} while sensing $J$ radar targets that \textit{may} also act as eavesdroppers (Eves), as shown in Fig.~\ref{fig:system_model}. 
The \ac{BS} can employ beamforming and \ac{AN} techniques to ensure secure communication while enabling target sensing, where each target direction $\theta_j$ of the $j$-th target is assumed to be pre-estimated, and secure transmission must be maintained against possible interception by these untrusted nodes (targets).

\subsection{Transmit and Receive Signal Models}

Under the described \ac{MU} \ac{ISAC} system, the transmit signal at the \ac{BS} is described by
\begin{equation}
\label{eq.transmit_signal}
\boldsymbol{x} = \boldsymbol{W}\boldsymbol{s} + \boldsymbol{n}_{\text{eff}} \in \mathbb{C}^{N_t \times 1},
\end{equation}  
where $\boldsymbol{s}=[s_1,\ldots,s_K]\trans \in \mathbb{C}^{K \times 1}$ is the data symbol vector for the $K$ \acp{CU}, $\boldsymbol{W}=[\boldsymbol{w}_1,\ldots,\boldsymbol{w}_K] \in \mathbb{C}^{N_t \times K}$ is the beamforming matrix with $\boldsymbol{w}_k \in \mathbb{C}^{N_t \times 1}$ denoting the $k$-th user's beamforming vector, and the injected \acf{AN} vector is given by $\boldsymbol{n}_{\text{eff}} \sim \mathcal{CN}(0,\boldsymbol{R}_{n_{\text{eff}}})$, which will serve to degrade the performance of the eavesdroppers while minimally affecting legitimate users and supporting sensing.

After transmission of the signal in \eqref{eq.transmit_signal} and the scattering reflection by the users and targets, the received echo signal at the \ac{BS} can be found, by defining $Q\triangleq K+J$ as the total number of reflectors, which include both the legitimate users and the unknown targets, as
\begin{align}
\boldsymbol{r} &= \sum_{q=1}^{Q} \alpha_q^2 \boldsymbol{a}(\theta_q) \boldsymbol{a}(\theta_q)^{\text{H}}\, \boldsymbol{x} + \boldsymbol{z} = \boldsymbol{G}(\boldsymbol{W}\boldsymbol{s} + \boldsymbol{n}_{\text{eff}}) + \boldsymbol{z}, 
\label{eq.received_signal} 
\end{align}
where $\alpha_q$ is the path gain, $\boldsymbol{z}\sim \mathcal{CN}(0,\sigma_r^2\boldsymbol{I})$ is the received noise at the \ac{BS}, and $\boldsymbol{a}(\theta_q)$ is the steering vector given by
\begin{equation}
\label{eq:steering_vector}
\boldsymbol{a}(\theta_q) = \big[1,\ e^{j2\pi \tfrac{d_M}{\lambda}\sin(\theta_q)},\ \ldots,\ e^{j2\pi \tfrac{d_M}{\lambda}(N_t-1)\sin(\theta_q)}\big]^{\text{T}}
\end{equation}
with inter-element spacing $d_M$ and wavelength $\lambda$.

\subsection{Secrecy Rate Formulation}

Given the described system and the signal models, the relevant performance metrics can be defined as follows.

First, we define the \ac{SINR} at legitimate user $k$ as
\begin{equation}
\label{eq.SINR_B_k}
\rho^{CU}_k = \frac{\boldsymbol{h}_k \boldsymbol{W}_k \boldsymbol{h}_k^{\text{H}}}{
\sum\limits_{i \ne k} \boldsymbol{h}_k \boldsymbol{W}_i \boldsymbol{h}_k^{\text{H}} +
\boldsymbol{h}_k \boldsymbol{R}_{n_{\text{eff}}} \boldsymbol{h}_k^{\text{H}} + \sigma_{z_k}^2},
\end{equation}
where $\boldsymbol{h}_k \in \mathbb{C}^{1 \times N_t}$ is the channel vector, $\boldsymbol{W}_k=\boldsymbol{w}_k\boldsymbol{w}_k^H$, and $\sigma_{z_k}^2$ is the receiver noise variance.

On the other hand, the intercepted \ac{SNR} at eavesdropper $j$ is given by
\begin{equation}
\label{eq.SINR_E_j}
\rho^E_{k,j} = 
\frac{|\alpha_j|^2 \boldsymbol{a}^{\text{H}}(\theta_j) \boldsymbol{W}_k \boldsymbol{a}(\theta_j)}{
|\alpha_j|^2 \boldsymbol{a}^{\text{H}}(\theta_j) \boldsymbol{R}_{n_{\text{eff}}} \boldsymbol{a}(\theta_j) + \sigma_e^2},
\end{equation}
where \(\sigma_e^2\) denotes the variance of the eavesdropper’s receiver noise, and all other parameters are defined according to the received echo signal model.

Given the above, the secrecy rate of user $k$ is given by
\begin{equation}
\label{eq.secrecy_rate}
\text{SR}_k = \left[\log_2(1+\rho^{CU}_k) - \max_j \log_2(1+\rho^E_{k,j})\right]^+,
\end{equation}
where $\rho^{CU}_k$ is the \ac{SINR} at user $k$, and $\rho^E_{k,j}$ is the received \ac{SNR} at eavesdropper $j$, which yields the secrecy sum rate as
\begin{equation}
\label{eq.sum_secrecy_rate}
\text{SR} = \sum_{k=1}^K \!\left[\log_2(1+\rho^{CU}_k) - \max_j \log_2(1+\rho^E_{k,j})\right]^+,
\end{equation}
where it can be seen that the above formulation inherently incorporates robustness by considering the worst-case eavesdropping scenario.

\section{Optimization Formulation}
\label{sec:structuring_opt}

In this section, we address the crucial design problem of the joint beamforming and \ac{AN} for secure \ac{ISAC}.  
Specifically, unlike conventional \ac{QoS}-driven formulations, we explicitly couple the communication and sensing requirements while constraining \ac{AN} to enhance radar returns without degrading legitimate user performance.
In the following, we study some constraints based on this requirement and formulate the final optimization problem.

First, to avoid interference with legitimate users, we assume \(K < N_T\)
and project the \ac{AN} vector onto the null space of the channel matrices of the users, i.e., 
\vspace{-1ex}
\begin{equation}
\boldsymbol{n}_{\text{eff}} = \boldsymbol{P}^\perp \boldsymbol{n}, \quad
\boldsymbol{P}^\perp = \boldsymbol{I}_{N_t} - \boldsymbol{H}^{\text{H}} (\boldsymbol{H}\boldsymbol{H}^{\text{H}})^{-1}\boldsymbol{H},
\vspace{-1ex}
\end{equation}
where the effective \ac{AN} covariance matrix becomes
\vspace{-1ex}
\begin{equation}
\boldsymbol{R}_{n_{\text{eff}}} = \boldsymbol{P}^\perp \boldsymbol{R}_n (\boldsymbol{P}^\perp)^{\text{H}}.
\vspace{-1ex}
\end{equation}

In addition, we also consider that the \ac{SINR} of each $k$-th user must remain above a minimum threshold given by $\gamma_k$, such that
\vspace{-1ex}
\begin{equation}
\rho_k^{CU} \geq \gamma_k, \quad \forall k.
\vspace{-1ex}
\end{equation}

On the other hand, to abide by the sensing requirements, we consider the radar received power at target $j$, given by
\begin{equation}
P_j = \alpha_j \boldsymbol{a}^{\text{H}}(\theta_j) 
\bigg(\sum\nolimits_{k=1}^K \boldsymbol{w}_k \boldsymbol{w}_k^{\text{H}} + \boldsymbol{R}_{n_{\text{eff}}}\bigg) 
\boldsymbol{a}(\theta_j).
\end{equation}

Following the above, the detection and beamwidth constraints are obtained as
\begin{eqnarray}
&\alpha_j \boldsymbol{a}^{\text{H}}(\theta_j) \tilde{\boldsymbol W}\boldsymbol{a}(\theta_j) \geq \eta_j, \quad \forall j,& \\
&\alpha_j \boldsymbol{a}^{\text{H}}(\theta_j \pm \theta_0)\tilde{\boldsymbol W}\boldsymbol{a}(\theta_j \pm \theta_0) \leq \tfrac{\eta_j}{2}, \quad \forall j,&
\end{eqnarray}
where $\eta_j$ is the predefined sensing power threshold, and $\theta_0$ is a specified angular interval threshold ensuring that 2$\theta_0$ is equal to the desired 3dB beamwidth of the waveform \cite{Li_2007}, ensuring sufficient angular resolution.

Then, considering the design variables and the constraints, the complete joint optimization problem is formulated as \vspace{1ex}
\vspace{-1ex}
\begin{align}
\label{eq.optimization_problem}
&\hspace{-6ex}\max_{\boldsymbol{w}_k, \boldsymbol{n}}  \sum_{k=1}^K \!\!
\left[ \!
\log_2 \!\! \Bigg(\!\!1 \!+\! 
\frac{\boldsymbol{h}_k \boldsymbol{w}_k \boldsymbol{w}_k^{\text{H}} \boldsymbol{h}_k^{\text{H}}}
{\sum\limits_{i \neq k} \! \boldsymbol{h}_k \boldsymbol{w}_i \boldsymbol{w}_i^{\text{H}} \boldsymbol{h}_k^{\text{H}} \!+\! 
\boldsymbol{h}_k \boldsymbol{n}_{\text{eff}} \boldsymbol{n}_{\text{eff}}^{\text{H}} \boldsymbol{h}_k^{\text{H}} \!+\! \sigma_k^2}
\!\!\Bigg) \right. \\[-2ex]
& \left.
\hspace{3ex} - \max_j \log_2 \left(1 + 
\frac{\alpha_j^2 \boldsymbol{a}^{\text{H}}(\theta_j) \boldsymbol{w}_k \boldsymbol{w}_k^{\text{H}} \boldsymbol{a}(\theta_j)}
{\alpha_j^2 \boldsymbol{a}^{\text{H}}(\theta_j) \boldsymbol{n}_{\text{eff}} \boldsymbol{n}_{\text{eff}}^{\text{H}} \boldsymbol{a}(\theta_j) + \sigma_e^2}
\right)
\right]^+ \!,\nonumber\\[2ex]
\text{s.t.} \quad 
&\nonumber
\text{(a)}\; \mathrm{Tr}(\boldsymbol{w}_k \boldsymbol{w}_k^{\text{H}}) \leq P_k, \quad \forall k, \\
\vspace{0.2em}
&\nonumber
\text{(b)}\; \mathrm{Tr}(\boldsymbol{R}_{n_{\text{eff}}}) + \sum_{k=1}^K P_k \leq P_A, \\
\vspace{0.2em}
&\nonumber
\text{(c)}\; \boldsymbol{a}^{\text{H}}(\theta_j) \tilde{\boldsymbol{W}} \boldsymbol{a}(\theta_j) \alpha_j \geq \eta_j, \quad \forall j, \\
\vspace{0.2em}
&\nonumber
\text{(d)}\; \boldsymbol{a}^{\text{H}}(\theta_j \pm \theta_0) \tilde{\boldsymbol{W}} \boldsymbol{a}(\theta_j \pm \theta_0) \alpha_j \leq \frac{\eta_j}{2}, \quad \forall j,\\
&\nonumber
\text{(e)}\; \rho^{CU}_k \geq \gamma_k, \quad \forall k. 
\end{align}

However, the above formulation includes non-smoothness and non-convexity introduced by the $[\,\cdot\,]^+$ operator and the $\max_j$, posing significant challenges for solving the problem.
Therefore, we reformulate the optimization problem by maximizing a weighted sum-rate of legitimate users subject to secrecy constraints that upper-bound the eavesdroppers' achievable rates as
\begin{align}
\label{eq.13}
&\hspace{-5ex} \max_{\boldsymbol{w}_k, \boldsymbol{n}_{\text{eff}}} \, \sum_{k=1}^K \mu_k\log_2\!\!\Bigg(\!\!1 \!+\! \frac {\boldsymbol{h}_k \boldsymbol{w}_k \boldsymbol{w}_k^{\text{H}} \boldsymbol{h}_k^{\text{H}}}
{\sum\limits_{i \neq k} \! \boldsymbol{h}_{k} \boldsymbol{w}_i \boldsymbol{w}_i^{\text{H}} \boldsymbol{h}_{k}^{\text{H}} \!+\! \boldsymbol{h}_k \boldsymbol{n}_{\text{eff}} \boldsymbol{n}_{\text{eff}}^{\text{H}} \boldsymbol{h}_k^{\text{H}} \!+\! \sigma_{k}^2}\!\! \Bigg) \\[-2ex]
\text{s.t.} \quad 
& \nonumber
\text{constraints 14(a)-14(d)}\\
\vspace{0.2em}
&\nonumber
\text{(f)}\;  \log_2\!\left(\!1 + \frac {\alpha_j^2 \boldsymbol{a}^{\text{H}}(\theta_j) \boldsymbol{w}_k \boldsymbol{w}_k^{\text{H}} \boldsymbol{a}(\theta_j)}{\alpha_j^2 \boldsymbol{a}^{\text{H}}(\theta_j) \boldsymbol{n}_{\text{eff}} \boldsymbol{n}_{\text{eff}}^{\text{H}} \boldsymbol{a}(\theta_j) + \sigma_e^2}\!\right) \!\leq\! \beta_j, \;\forall j
\vspace{0.2em}
\end{align}
which has now been converted into a more tractable form, enabling the use of standard optimization techniques while ensuring a predefined level of secrecy.

The parameter \(\mu_k\) is introduced to allow the base station to adjust user weights if the final SNRs fail to meet the constraint \( \rho^{CU}_k \geq \gamma_k\). This parameter can either be optimized during the process or leveraged to enforce the fairness condition.

\section{Dual Optimization Framework}
\label{sec:dual_opt_frmework}

In this section, we motivate that the reformulated problem in \eqref{eq.13} can be decomposed into two iterative optimization subproblems: optimizing the beamforming vectors $\boldsymbol{w}_k$, and optimizing the artificial noise $\boldsymbol{n}$.
Accordingly, we develop a dual optimization framework to address the joint beamforming and noise design problem, leveraging a dual \acf{QT} approach.

\subsection{Subproblem I: Non-homogeneous \ac{QT} for Beamforming}

We first optimize the beamforming vectors $\boldsymbol{w}_k$ with $\boldsymbol{n}$ fixed. In this step, only constraints (a)–(b) in \eqref{eq.13} are enforced, while the sensing constraints are deferred to Subproblem~II.  
Then, the equivalent first optimization problem can be expressed as
\vspace{-1ex}
\begin{align}
\label{eq.14}
\max _{\boldsymbol{w}_k} \quad & \sum_{k=1}^K \mu_k\log_2\left(1 + M_k(\boldsymbol{w}_k) \right) \\
\text{s.t.}\; \quad&
\text{constraints 14(a) and 15(f)},\nonumber\\[-4ex]
\end{align}
\noindent where 
\begin{subequations}
\label{eq:all}
\begin{eqnarray}
&M_k(\boldsymbol{w}_k) = e_k^*(\boldsymbol{w}_k) B_k^{-1}(\boldsymbol{w}_k) e_k(\boldsymbol{w}_k),&\label{eq:Mk} \\
&e_k(\boldsymbol{w}_k) = \boldsymbol{h}_k \boldsymbol{w}_k,& \\
&B_k(\boldsymbol{w}_k) = \sum_{i \neq k} \boldsymbol{h}_k \boldsymbol{w}_i \boldsymbol{w}_i^H \boldsymbol{h}_k^H 
+ \boldsymbol{h}_k \boldsymbol{R}_{n_{\text{eff}}} \boldsymbol{h}_k^H + \sigma_k^2.&
\end{eqnarray}
\end{subequations}

Next, observe that constraint 15(f) can be expressed as a convex inequality if $\boldsymbol{n}$ is fixed, while the objective function of \eqref{eq.14} can be reformulated as \cite{ShenTSP2018, II-ShenTSP2018}
\begin{equation}
\label{eq.21}
h(\boldsymbol{w}_k,\boldsymbol{\zeta}) \!=\! \sum_{k=1}^{K} \mu_k \! \left[(1\!+\!\zeta_k)\hat{M}_k(\boldsymbol{w}_k) \!+\! \log_2(1\!+\!\zeta_k) \!-\! \zeta_k \right],
\end{equation}
where $\bm{\zeta} \triangleq \{\zeta_1, \cdots, \zeta_K\}$ denotes the set of auxiliary variables, and the implicitly defined updated variables are given by
\begin{subequations}
\begin{align}
\hat{M}_k(\boldsymbol{w}_k) &\triangleq e^*_k (\boldsymbol{w}_k) \hat{B}_k^{-1}(\boldsymbol{w}_k) e_k(\boldsymbol{w}_k), \label{eq:Mk2}\\[1ex]
\hat{B}_k(\boldsymbol{w}_k) & \triangleq e_k(\boldsymbol{w}_k) e^* _k(\boldsymbol{w}_k) + B_k(\boldsymbol{w}_k) \label{eq:Bk2}.
\end{align}
\end{subequations}

Next, by utilizing a dual complex \ac{QT}, we can re-express $\hat{M}_k(\boldsymbol{w}_k)$ as 
\begin{equation}
\label{eq.20}
\hat{M}_k(\boldsymbol{w}_k) = 2{\operatorname{Re}}\left\{ y_k^* e_k (\boldsymbol{w}_k)\right\} -y_k^* \hat{B}_k(\boldsymbol{w}_k)y_k,
\end{equation}
where $y_k \in \mathbb{C}$ is an auxiliary variable.

To solve the problem efficiently, we adopt the non-homogeneous quadratic transform, a recent advancement in \ac{FP} shown to outperform the classical \ac{QT} in both convergence and efficiency \cite{ShenJSAC2024,Sun_TCP2017}. 

This leads to the equivalent quadratic dual for the optimization objective in \eqref{eq.21}, given by 
\begin{eqnarray}
\label{eq.26}
f_k(\boldsymbol{w}_k, y_k, \boldsymbol{z}_k, \zeta_k) = \sum_{k=1}^{K} \mu_k \left[ \log_2(1+\zeta_k) - \zeta_k \right] +&&\\[-1ex]
&&\hspace{-48ex}
\sum_{k=1}^{K} \! \bigg(
\!2 \operatorname{Re} \! \bigg\{ \mu_k(1+\zeta_k)y_k^* \boldsymbol{h}_k \boldsymbol{w}_k + \boldsymbol{w}_k^{\text{H}}\left( \kappa_k \boldsymbol{I}_{N_t} \!-\! \boldsymbol{D}_k\right)\boldsymbol{z}_k \!\bigg\}\! 
\nonumber \\[-1ex]
&&\hspace{-45ex}+ \boldsymbol{z}_k^{\text{H}} \left( \boldsymbol{D}_k - \kappa_k \boldsymbol{I}_{N_t}\right)\boldsymbol{z}_k  - \kappa_k \boldsymbol{w}_k^{\text{H}} \boldsymbol{w}_k - \mu_k(1+\zeta_k)\sigma_k^2 y_k^* y_k\nonumber \\
&&\hspace{-45ex}- \mu_k(1+\zeta_k)\boldsymbol{n}^{\text{H}} (\boldsymbol{P}^\perp)^{\text{H}} \boldsymbol{h}_k^{\text{H}} y_k y_k^* \boldsymbol{h}_k \boldsymbol{P}^\perp \boldsymbol{n} \bigg)  .\nonumber
\end{eqnarray}

From the above, the closed-form optimum of the updated beamforming weights \({\boldsymbol{w}}_{k}^{\star}\) can be obtained as
\begin{equation}
\label{eq.31}
{\boldsymbol{w}}_{k}^{\star} = \boldsymbol{z}_{k}^{\star} + \frac{1}{\kappa_{k}}\left(\mu_{k}\left(1+\zeta_{k}^{\star}\right) \boldsymbol{h}_{k}^{\mathrm{H}} y_{k}^{\star} -\boldsymbol{D}_{k} \boldsymbol{z}_{k}^{\star} \right)
\end{equation}
with
\vspace{-1ex}
\begin{subequations}
\begin{eqnarray}
\label{eq.29}
&y_k^{\star} = \hat{B}_k^{-1}(\boldsymbol{w}_k) e_k(\boldsymbol{w}_k),&\\
\label{eq.30}
&\zeta_k^* = \dfrac{1}{\ln(2) \left( 1\! +\! \tilde{n}_k \!+\! \sigma_k^2 y_k y_k^* \!- \!\operatorname{Re}  \{y_k e_k(\boldsymbol{w}_k) \}  \right)} \!- \!1,&
\end{eqnarray}
initialized by $\boldsymbol{z}_k^{\star} = \boldsymbol{w}_k$.
\end{subequations}

Based on the closed-form solutions, the beamforming weights are updated at the $n$-th iteration as
\begin{equation}
\label{eq.32}
\!\!\!\hat{\boldsymbol{w}}_{k}^{(n)}\! =\! \mathcal{P}_{\boldsymbol{w}_{k} \in \mathcal{W}_{k}} \!\!\left(\!\boldsymbol{z}_{k}^{\star} \!+ \!\frac{1}{\kappa_{k}}\!\left(\mu_{k}\left(1+\zeta_{k}^{\star}\right) \boldsymbol{h}_{k}^{\mathrm{H}} y_{k}^{\star} -\boldsymbol{D}_{k} \boldsymbol{z}_{k}^{\star} \right)\!\right)\!,
\!\!\!
\end{equation}
where \(\mathcal{P}_{\boldsymbol{w}_{k} \in \mathcal{W}_{k}}\) denotes projection onto the feasible set \(\mathcal{W}_{k}\), and $\kappa_k$ is chosen to upper-bound the largest eigenvalue of $\boldsymbol{D}_k$.

Thus, Subproblem~I reduces to an iterative algorithm with simple closed-form updates for $y_k$, $\zeta_k$, and $\boldsymbol{w}_k$, avoiding costly matrix inversions while ensuring convergence.

\subsection{Subproblem II: Non-homogeneous \ac{QT} for \ac{AN}}

In turn, we optimize over the artificial noise vector $\boldsymbol{n}$ while fixing $\{\boldsymbol{w}_k\}$ in this section, whose objective is given by
\begin{align}
\label{eq.33_short}
&\hspace{-5ex}\max _{\boldsymbol{n}} \,\sum_{k=1}^K\! \mu_k\log_2\!\bigg(1 \!+ \!\frac{|\boldsymbol{h}_k\boldsymbol{w}_k|^2}{\sum_{i \neq k}|\boldsymbol{h}_k\boldsymbol{w}_i|^2\! +\! \boldsymbol{h}_k\boldsymbol{R}_{n_{\text{eff}}}\boldsymbol{h}_k^H \!+\! \sigma_k^2}\bigg) \\
\text{s.t.} \;
&\text{constraints 14(c-d)},\nonumber\\[-2ex]
&\text{(g)}\; \mathrm{Tr}(\boldsymbol{P}^\perp \boldsymbol{n}\boldsymbol{n}^H(\boldsymbol{P}^\perp)^H) + \sum_{k=1}^K P_k \leq P_A,\nonumber
\end{align}

Leveraging a procedure similar to the previous subproblem, the objective function with respect to $\boldsymbol{n}$ can be equivalently stated by introducing auxiliary variables $\tilde{\bm{\zeta}}=\{\tilde{\zeta}_k\}$ and $\tilde{y}_k$ as
\begin{align} 
\label{eq.36}
\!\!h(\boldsymbol{n},\tilde{\bm{\zeta}}) \!=\! \sum^K_{k=1}\tilde{\mu}_k \bigg[ \!\log_2 \,(1+\tilde{\zeta}_k ) \!-\!  \tilde{\zeta}_k + (1+\tilde{\zeta}_k)\hat{M}_k(\boldsymbol{n}) \bigg],\!
\end{align}
where
\begin{subequations}
\begin{eqnarray}
&\hat{M}_k(\boldsymbol{n}) \triangleq e^*_k (e_k e^* _k + B_k(\boldsymbol{n}))^{-1}e_k,&\\
&\hat{B}_k(\boldsymbol{n}) \triangleq e_k e^*_k + B_k(\boldsymbol{n}),&
\end{eqnarray}
\end{subequations}
i.e., by letting $M_k(\boldsymbol{n})$ to be identical to $M_k(\boldsymbol{w}_k)$ in \eqref{eq:Mk} with the emphasis on $\boldsymbol{n}$, and we drop the inherent dependence of $\boldsymbol{w}_k$ on $e_k(\boldsymbol{w}_k)$ and hereafter denote it as $e_k$.

Trivially, it can be seen that the reformulated objective $f_r(\cdot)$ follows the non-homogeneous \ac{QT} structure as in Subproblem~I, and therefore similar derivations can lead to the closed-form update given by
\begin{equation}
\boldsymbol{n}^\star = \left(\sum_{k=1}^K \tilde{\kappa}_k \boldsymbol{I}_{N_t}\right)^{-1} \sum_{k=1}^K (\tilde{\kappa}_k \boldsymbol{I}_{N_t} - \tilde{\boldsymbol{D}}_k)\,\boldsymbol{n}, \label{eq:n_update},
\end{equation}
with
\begin{subequations}
\begin{equation}
\tilde{y}_k^\star = \hat{B}_k^{-1}(\boldsymbol{n}) e_k, \label{eq:y_update}
\vspace{-2ex}
\end{equation}
\begin{align}
\tilde{\zeta}_k^\star &= \frac{1}{\ln(2)} \!\bigg[ \boldsymbol{w}_k^H\!\bigg(\boldsymbol{h}_k^H\tilde{y}_k\tilde{y}_k^*\boldsymbol{h}_k + \!\!\sum_{i\neq k}\!\boldsymbol{h}_k^H\tilde{y}_i\tilde{y}_i^*\boldsymbol{h}_k\bigg)\!\boldsymbol{w}_k  \label{eq:zeta_update} \\[-2ex]
&\hspace{18ex}+ \sigma_k^2|\tilde{y}_k|^2 - 2{\operatorname{Re}}\{\tilde{y}_k^*\boldsymbol{h}_k\boldsymbol{w}_k\}\!\bigg]^{\!-1} \!\!\!\!-\!1, \nonumber
\end{align}
and $\tilde{\kappa}_k$ again upper-bounds the largest eigenvalue of $\tilde{\boldsymbol{D}}_k$.
\end{subequations}

Then, at iteration $t$, the updated value is projected onto the feasible set as
\begin{equation}
\hat{\boldsymbol{n}}^{(t)} \!=\! \mathcal{P}_{\boldsymbol{n}\in\mathcal{N}}\Bigg( \!\bigg(\sum_{k=1}^K \tilde{\kappa}_k \boldsymbol{I}_{N_t}\!\bigg)^{\!\!-1}\!\sum_{k=1}^K (\tilde{\kappa}_k \boldsymbol{I}_{N_t}\!\!-\!\tilde{\boldsymbol{D}}_k)\hat{\boldsymbol{n}}^{(t-1)}\!\Bigg),\!
\label{eq:finaln}
\end{equation}
where \(\mathcal{P}_{\boldsymbol{n} \in \mathcal{N}}\) denotes the projection onto the feasible set \(\mathcal{N}\).

\begin{algorithm}[H]
\caption{Proposed Solution to the Secure ISAC Problem}
\label{alg:NQT}
\begin{algorithmic}[1]
\STATE \textbf{Input:} Vector \(\boldsymbol{\mu}\)
\STATE Initialize $\boldsymbol{w}$ and $\boldsymbol{n}$.
\REPEAT
\STATE \textbf{Optimization Problem 1} ($\boldsymbol{n}$ is held fixed)
\FOR{each $k$}
\STATE Initialize auxiliary variable $\boldsymbol{z}_k$ to $\boldsymbol{w}_{k}$.
\STATE Update auxiliary variable $y_k$ via \eqref{eq.29}.
\STATE Update auxiliary variable $\tilde{\zeta}_k$ via \eqref{eq.30}.
\STATE Update projected weight variable $\boldsymbol{w}_k$ via \eqref{eq.32}.

\ENDFOR

\STATE \textbf{Optimization Problem 2} ($\boldsymbol w_k$ is held fixed)
\STATE Initialize auxiliary variable $\boldsymbol{z}_k$ to $\boldsymbol{n}$.
\STATE Update auxiliary variable $\tilde{y}_k$ via \eqref{eq:y_update}.
\STATE Update auxiliary variable $\tilde{\zeta}_k$ via \eqref{eq:zeta_update}.
\STATE Update projected noise variable $\boldsymbol{n}$ via \eqref{eq:finaln}.

\UNTIL{the objective function in \eqref{eq.13} converges}
\end{algorithmic}
\end{algorithm}
\vspace{-2ex}

Thus, similarly to Subproblem~I, Subproblem~II is solved by iteratively updating $\tilde{y}_k$, $\tilde{\zeta}_k$, and $\boldsymbol{n}$ in closed form, followed by a projection to the valid set of the \ac{AN} variables.

In all, the complete steps for the proposed approach, including both subproblems, are summarized as pseudocode with reference to the update rules in Algorithm \ref{alg:NQT} below.

\section{Robust Beamforming under Angular Uncertainty}

To account for angle estimation errors, the $j$-th target angle is modeled as
\begin{equation}
\theta_j \in [\hat{\theta}_j - \Delta_j, \hat{\theta}_j + \Delta_j],
\end{equation}
where $\hat{\theta}_j$ is the nominal DoA and $\Delta_j$ the maximum deviation. The worst-case eavesdropper SNR is
\begin{equation}
\rho^E_{k,j}(\theta_j) =
\frac{|\alpha_j|^2 \boldsymbol{a}^{\text{H}}(\theta_j)\boldsymbol{W}_k\boldsymbol{a}(\theta_j)}
{|\alpha_j|^2 \boldsymbol{a}^{\text{H}}(\theta_j)\boldsymbol{R}_{n_{\text{eff}}}\boldsymbol{a}(\theta_j)+\sigma_e^2},
\end{equation}
and the robust secrecy rate is
\vspace{-2ex}
\begin{equation}
\text{SR}_k^{\text{robust}} = \Bigg[\log_2(1+\rho^{CU}_k) - 
\max_{j, \theta_j \in [\hat{\theta}_j \pm \Delta_j]} \log_2(1+\rho^E_{k,j}(\theta_j)) \Bigg]^+.
\end{equation}

For small $\Delta_j$, the steering vector can be linearized as
\begin{equation}
\boldsymbol{a}(\theta_j) \approx \boldsymbol{a}(\hat{\theta}_j) + (\theta_j - \hat{\theta}_j)\boldsymbol{a}'(\hat{\theta}_j),
\end{equation}
so that the worst-case eavesdropper SNR reduces to
\begin{equation}
\max_{\theta_j \in [\hat{\theta}_j \pm \Delta_j]} \rho^E_{k,j}(\theta_j) 
\approx \max\{\rho^E_{k,j}(\hat{\theta}_j - \Delta_j), \rho^E_{k,j}(\hat{\theta}_j + \Delta_j)\}.
\end{equation}

\vspace{-2ex}
The robust beamforming problem is formulated as
\vspace{-2ex}
\begin{align}
\max_{\boldsymbol{w}_k, \boldsymbol{n}} & \quad \sum_{k=1}^{K} \text{SR}_k^{\text{robust}} \\
\text{s.t.} & \quad \eqref{eq.optimization_problem}\: \text{a-d,} \text{\:considering\:} \theta_j = \hat{\theta}_j \pm \Delta_j. \nonumber
\end{align}
By defining the robust upper bound for each eavesdropper as
\vspace{-1ex}
\begin{equation}
\beta_{k,j} = \max_{\theta_j = \hat{\theta}_j \pm \Delta_j} 
\frac{|\alpha_j|^2 \boldsymbol{a}^{\text{H}}(\theta_j)\boldsymbol{W}_k\boldsymbol{a}(\theta_j)}
{|\alpha_j|^2 \boldsymbol{a}^{\text{H}}(\theta_j)\boldsymbol{R}_{n_{\text{eff}}}\boldsymbol{a}(\theta_j)+\sigma_e^2},
\end{equation}
the S-procedure allows the corresponding LMI constraint
\begin{equation}
\boldsymbol{W}_k \preceq \frac{\beta_{k,j}}{|\alpha_j|^2}
\Big(\boldsymbol{a}(\hat{\theta}_j)\boldsymbol{a}^{\text{H}}(\hat{\theta}_j) +
\Delta_j^2 \boldsymbol{a}'(\hat{\theta}_j)\boldsymbol{a}'(\hat{\theta}_j)^{\text{H}} \Big)^{-1}.
\end{equation}

The dual quadratic transform framework can then be applied to iteratively update the beamforming and artificial noise matrices while satisfying the robust constraints. The beamforming vectors are updated as
\begin{equation}
\hat{\boldsymbol{w}}_k^{(n)} = \mathcal{P}_{\mathcal{W}_k^{\text{robust}}} 
\Big( \boldsymbol{z}_k^\star + \tfrac{1}{\kappa_k} 
(\mu_k(1+\zeta_k^\star)\boldsymbol{h}_k^H y_k^\star - \boldsymbol{D}_k\boldsymbol{z}_k^\star)\Big),
\end{equation}
while the artificial noise matrix is updated according to
\vspace{-2ex}
\begin{equation}
\hat{\boldsymbol{n}}^{(t)} = \mathcal{P}_{\mathcal{N}^{\text{robust}}} 
\Bigg( \bigg( \sum_{k=1}^K \tilde{\kappa}_k \boldsymbol{I}_{N_t} \bigg)^{-1} 
\sum_{k=1}^K (\tilde{\kappa}_k \boldsymbol{I}_{N_t} - \tilde{\boldsymbol{D}}_k)\hat{\boldsymbol{n}}^{(t-1)} \Bigg),
\end{equation}
where the feasible sets $\mathcal{W}_k^{\text{robust}}$ and $\mathcal{N}^{\text{robust}}$ incorporate the \ac{LMI} constraints to guarantee worst-case secrecy under angular uncertainty.

\section{Performance Analysis}
\label{sec:performance_analysis}

This section evaluates the performance of the proposed framework through Monte Carlo simulations.
The results highlight the effectiveness of the joint beamforming and artificial noise design. 
Unless otherwise stated, the channel coefficients $\boldsymbol{h}_k$ are modeled as i.i.d. complex Gaussian random variables with distribution $\mathcal{CN}(0,1)$.
The \ac{ISAC} base station employs a \ac{ULA} with half-wavelength antenna spacing, and the noise variance at the targets is assumed equal to that of the intended \acp{CU}. 
The key simulation parameters are summarized in Table~\ref{tab:simulation_parameters}.

\begin{figure}[H]
\centering
\includegraphics[width=0.9\columnwidth]{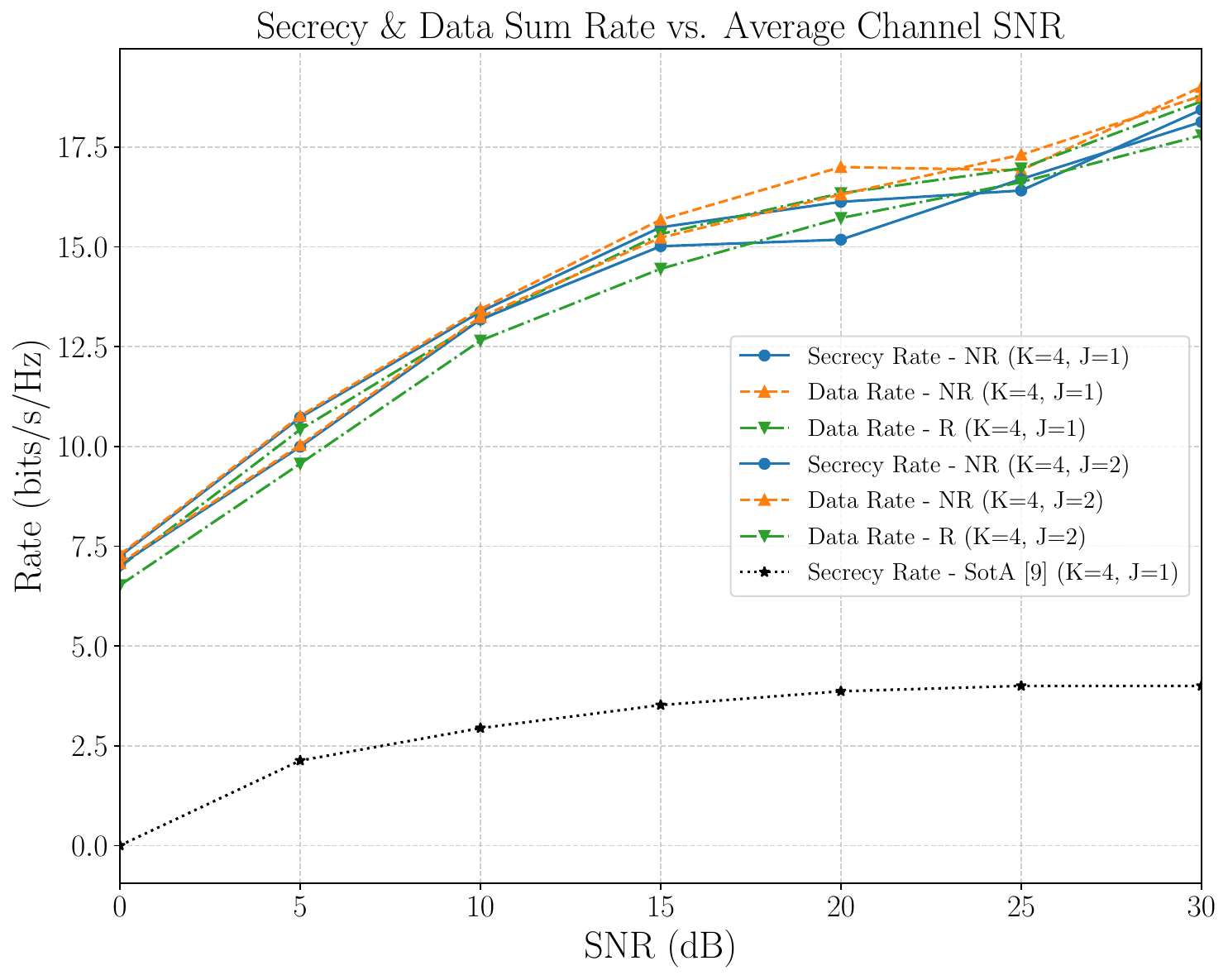}
\vspace{-2ex}
\caption{Average secrecy rate performance of the proposed Algorithm~\ref{alg:NQT} under varying numbers of users $K$, targets $J$, and transmit antennas $N_t$, compared with a representative \ac{SotA} method.}
\label{fig:secrecyRate}
\end{figure}

Fig. \ref{fig:secrecyRate} shows the average secrecy rate as a function of the number of users and eavesdroppers.
Larger values of $J$ reduce performance, as more interference and eavesdropping threats must be mitigated.
Compared with the state-of-the-art benchmark in \cite{SuTWC2021}, the proposed method achieves consistently higher secrecy rates across all \ac{SNR} values.
In particular, while the benchmark method saturates at an average secrecy rate of roughly 4~bit/s/Hz, our approach continues to scale with system size.

\begin{figure}[H]
\centering
\includegraphics[width=0.9\columnwidth]{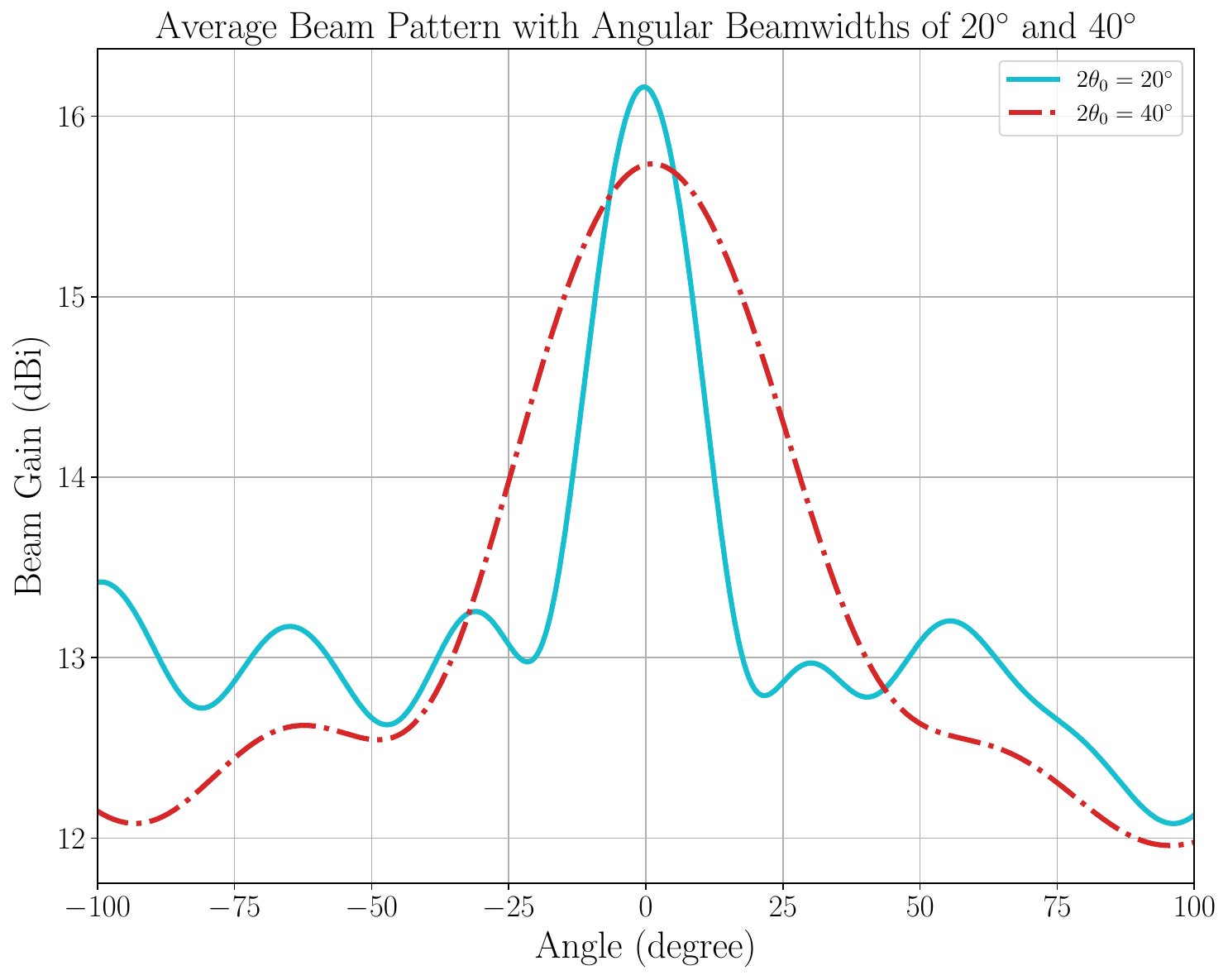}
\vspace{-2ex}
\caption{Beam gain of the proposed Algorithm~\ref{alg:NQT} for various target directions with $N_t=16$, $K=4$, and $J=1$.}
\label{fig:beamGain}
\end{figure}

The sensing performance is illustrated in Fig.~\ref{fig:beamGain}, which plots the beam gain across different eavesdropper directions. 
The SINR threshold for each legitimate user is fixed at 20~dB. 
For comparison, a narrow beam pattern assuming perfect target direction knowledge at the base station is also shown. 
\vspace{-3ex}
\begin{table}[H]
\centering
\caption{Simulation Parameters for Joint Optimization}
\label{tab:simulation_parameters}
\setlength{\tabcolsep}{4pt} 
\renewcommand{\arraystretch}{1} 
\begin{tabular}{|c|c|c|}
\hline
Parameter & Symbol & Value \\
\hline
Number of users & $K$ & 2, 4, 8 \\
Number of eavesdroppers & $J$ & 1, 2 \\
Transmit antennas & $N_T$ & 8, 16, 18 \\
Total power & $P_A$ & 20 dBm \\
Eavesdropper angles & $\theta_j$ & $[-90^\circ, +90^\circ]$ \\
Mainlobe leakage & $\theta_0$ & $[0^\circ, 20^\circ]$ \\
Secrecy SINR threshold & $\beta_j$ & 0.1  \\
Minimum sensing gain & $\eta_j$ & 2 \\
Path loss (eavesdropper) & $\alpha_j$ & 1 \\
\hline
\end{tabular}
\end{table}

While the reference method yields a sharp peak, the proposed algorithm produces wider main lobes, sustaining high gain across the entire angular span of $[-90^\circ, +90^\circ]$. 
Although the gain decreases as the region of interest expands, reliable sensing is maintained even for $\theta_0 = 20^\circ$, confirming its robustness.

\begin{figure}[h]
\centering
\includegraphics[width=0.9\columnwidth]{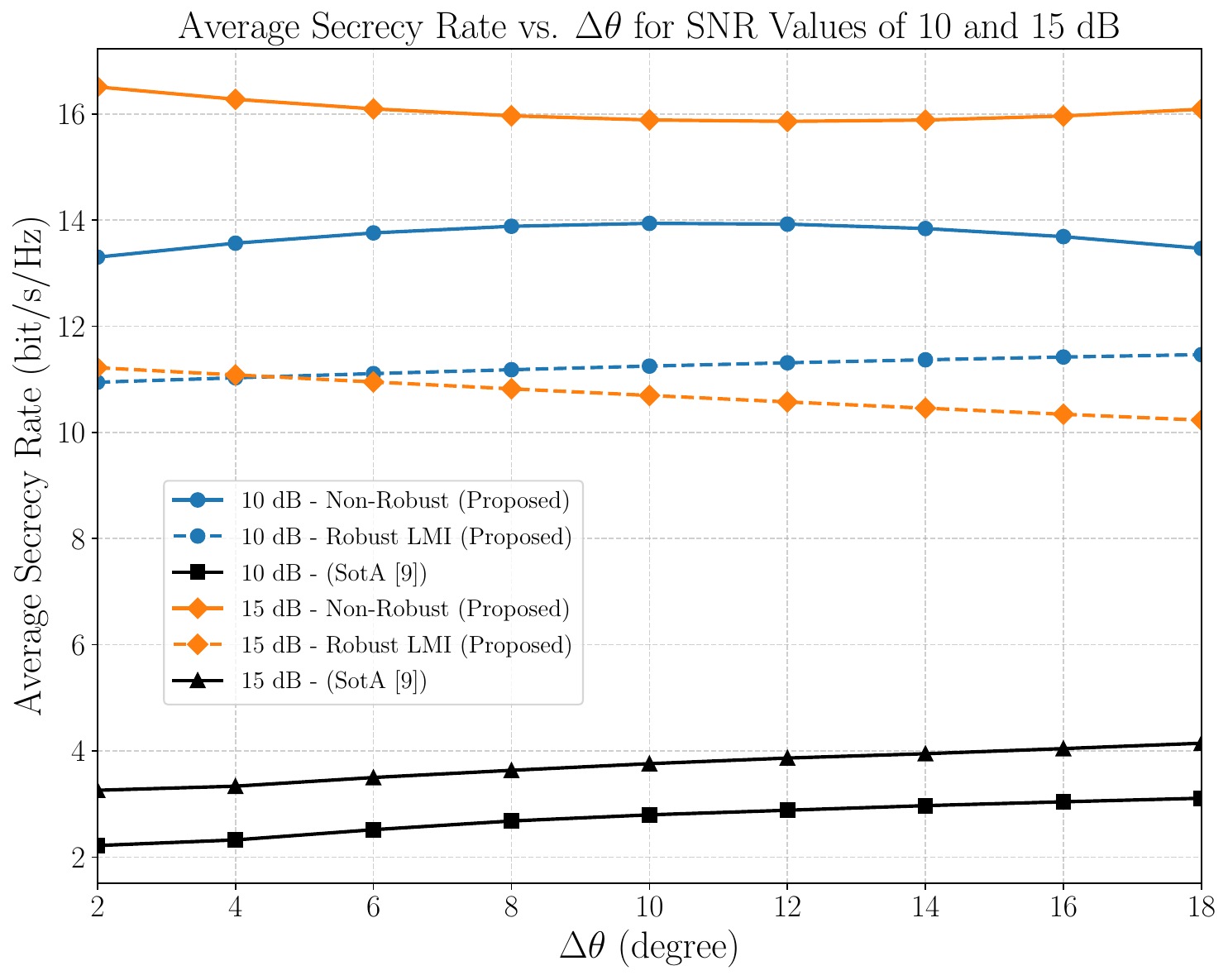}
\vspace{-2ex}
\caption{Average secrecy rate versus angular variation $\Delta \theta$ for the proposed Algorithm~\ref{alg:NQT} with $N_t=18$, $K=4$, and $J=1$.}
\label{fig:deltaTheta}
\end{figure}

Finally, Fig.~\ref{fig:deltaTheta} further investigates the impact of angular variation on secrecy rate.
Here, the proposed algorithm consistently outperforms the \ac{SotA} across all $\Delta \theta$ values, sustaining higher secrecy rates throughout. 
Importantly, due to the artificial noise design, performance remains stable even as $\Delta \theta$ grows, in contrast to the pronounced fluctuations exhibited by the benchmark method.

\section{Conclusion}
\label{sec:conclusion}
\vspace{-0.3em}
We introduced a secure \ac{MU}-\ac{ISAC} framework that jointly addresses communication, sensing, and fairness.
By formulating secrecy rate maximization with adversarial targets, we developed an efficient dual fractional programming approach using a non-homogeneous quadratic transform. 
The proposed approach enables closed-form updates for both beamforming and artificial noise, ensuring scalability to large \ac{MIMO} systems. 
Simulation results demonstrated clear gains over existing schemes in secrecy rate and beamforming robustness, highlighting the framework's suitability for next-generation secure \ac{ISAC} networks.

\vspace{-2ex}
\bibliographystyle{ieeetr}

\end{document}